\documentclass{emulateapj}
\usepackage{times}
\usepackage{amsmath}
\usepackage{graphicx}

\slugcomment{Submitted to ApJ }

\shorttitle{PSR J2124$-$3358 Bowshock}
\shortauthors{Romani, Slane \& Green }

\begin{document}

\title{The Asymmetric Bowshock/PWN of PSR J2124$-$3358}

\author{Roger W. Romani \altaffilmark{1}, Patrick Slane \& Andrew Green}
\altaffiltext{1}{Department of Physics, Stanford University, Stanford, CA 94305-4060,
 USA; rwr@astro.stanford.edu}

\begin{abstract}

	We describe new measurements of the remarkable H$\alpha$/UV/X-ray
bowshock and pulsar wind nebula of the isolated millisecond pulsar PSR J2124$-$3358.
{\it CXO} imaging shows a one-sided jet structure with a softer equatorial
outflow. KOALA IFU spectroscopy shows that non-radiative emission dominates
the bow shock and that the H$\alpha$ nebula is asymmetric about the pulsar
velocity with an elongation into the plane of the sky. We extend analytic
models of the contact discontinuity to accommodate such shapes and compare
these to the data. Using {\it HST} UV detections of the pulsar and bow shock, 
radio timing distance and proper motion measurements and the {\it CXO}-detected 
projected spin axis we model the 3-D PWN momentum flux distribution. The
integrated momentum flux depends on the ionization of the ambient ISM, but for
an expected ambient WNM we infer $I = 2.4 \times 10^{45} {\rm g\,cm^2}$.
This implies $M_{NS} = 1.6-2.1 M_\odot$, depending on the equation of state,
which in turn suggests that the MSP gained significant mass during recycling and 
then lost its companion. However, this conclusion is at present tentative,
since lower ionization allows $\sim 30\%$ lower masses and uncertainty in
the parallax allows up to 50\% error.
\end{abstract}

\keywords{pulsars: individual: PSR J2124$-$3358 -- shock waves -- dense matter}

\section{Introduction}

	The standard paradigm for pulsar recycling involves mass and angular
momentum accretion in a binary \citep{bkk74,sb76}. Yet a number of recycled
pulsars, especially short period millisecond pulsars (MSP) are isolated.
An attractive scenario for such single MSP invokes evolution through an extreme
`Black Widow' phase where the MSP spindown power completely evaporates a companion \citep{rst89}.
The critical prediction of this scenario is a heavy neutron star.
For an isolated (non-binary) pulsar this would seem impossible to test. 
However, we have proposed \citep[hereafter BR14]{br14} that if a sufficiently accurate measurement
of the total spindown power ${\dot E}$ can be made, using detailed measurements
of a pulsar bow shock, this can be compared with $I \Omega {\dot \Omega}$,
yielding the neutron star moment of inertia $I$. Since $I$ increases along
with mass, this provides a test of the recycling scenario.

	The H$\alpha$ bow shock of PSR J2124$-$3358 (hereafter J2124) discovered by \citet{gjs02}
provides an opportunity for such a test. This 4.9\,ms pulsar is nearby and
pulsar timing has provided a high accuracy parallax distance $d=0.41^{+0.09}_{-0.07}$\,kpc
and proper motion $v_\perp=101.2\pm0.8 {\rm km\,s^{-1}}$ at
a position angle of 195.77$\pm0.08^\circ$ \citep{reet17}. If we correct for the standard
solar motion \citep{js87} at $d=410$\,pc the transverse velocity is $v_\perp=110.1\pm0.8 
{\rm km\,s^{-1}}$ at 202.98$\pm0.08^\circ$.  BR14 made new,
deeper H$\alpha$ images of the nebula, measuring the shape and flux of the apex.
An archival December 19, 2004 30\,ks {\it CXO} ACIS observation of the pulsar (ObsID 5585, Chatterjee, PI) 
revealed, in addition to the pulsar point source, a diffuse X-ray pulsar wind nebula (PWN) 
within the H$\alpha$ bowshock with a long jet or trail extending to the NW, 
following the H$\alpha$ limb. These data were briefly analyzed by \citet{hb06},
who also studied a lower-resolution {\it XMM-Newton} exposure. Their analysis
showed a soft point source and somewhat harder spectrum for the PWN jet/trail.

	Recently \citet{raet17} have reported on {\it HST SBC/WFC3} observations of
the pulsar. These data detect both the pulsar point source and an arc of UV
emission (best seen in the F125LP filter) from the bow shock itself. These data are
very important in measuring the distance from the pulsar to the shock limb which
is $1.8^{\prime\prime}$ along the projected pulsar velocity vector. We show below
that, within ground-based resolution, this is also the offset to the $H\alpha$ shock limb.
This very small stand-off, and the flat shock near the apex are important
constraints on the pulsar wind distribution and integrated momentum flux. In this paper we explore 
the PWN geometry with new observations and models. These give
insights into the pulsar wind structure and spindown
energetics. We comment on the implications for the neutron star moment of inertia
and the recycling scenario.

\section{New Observations}

	We describe here new, deeper {\it CXO} observations probing the PWN spectrum
and morphology, an attempt at improved  {\it HST} astrometry and a new AAT/KOALA
observation to measure the velocity structure and optical spectrum of the bow shock
apex. Together these data give a refined view of the PWN/bow shock structure,
and conditions in the surrounding ISM.

\subsection{Chandra imaging}

	We obtained new {\it CXO} ACIS-S imaging of J2214 with two exposures:
93\,ks on July 7, 2016 (ObsID 17900) and 85\,ks on September 4, 2016 (ObsID 19686)
with the source placed near the aimpoint of the S3 chip.
All data were useful with no strong background flares. Standard
ACIS processing was used to align all exposures to the 
December 19, 2004 (ObsID 5585) archival frame, checked
by matching field sources. The $0.6^{\prime\prime}$ proper motion shift over
the 11.5 year gap was obvious.  The overall spectrum is very soft, with only a few
point source counts above 3\,keV. This is not unexpected, given the close distance.
This distance and the low dispersion measure $DM=4.6 {\rm cm^{-3} pc}$ imply an absorption
column $N_H\approx 1-3 \times 10^{20} {\rm cm^{-2}}$ \citep{raet17}. With such low absorption
and a relatively soft source, the progressive ACIS contamination 
and low energy effective area loss are particularly severe. However, with the longer 2016 exposures, 
we do accrue somewhat larger image counts than in the the archival observation.

	We re-aligned the data, referencing to the point source (2004.97 epoch) and merged
the exposures. The improved statistics provide a better view of fine structure
near the pulsar. The strongest feature is the extension that leaves the pulsar
at $PA = -60^\circ$, and then sweeps north to stay within the H$\alpha$ limb. 
This is most prominent at intermediate $\sim 1.5-3\,$keV energies. We will refer to this
as the `Trail' although as argued below, its morphology and relatively hard spectrum
suggest identification as a pulsar polar jet.  At lower $<1.5\,$keV energy, an elliptical extension
to the point source is apparent, with major axis radius $\sim 3^{\prime\prime}$ 
transverse to the jet. We identify this softer emission with an equatorial outflow
describable as a `torus' in many PWNe. To the NE ($PA\approx 45^\circ$) this
emission persists for another $\sim 4^{\prime\prime}$. This can be interpreted as
the sweep back of the equatorial outflow, as it reverse shocks inside the 
H$\alpha$ forward shock envelope. Although there are a handful of X-ray counts
along the axis opposite the jet, such a `forward' jet is not well detected. 
Similarly the equatorial outflow appears to be terminated to the SW, where
it reaches the forward shock.

\begin{figure}[t!!]
\vskip 6.4truecm
\includegraphics{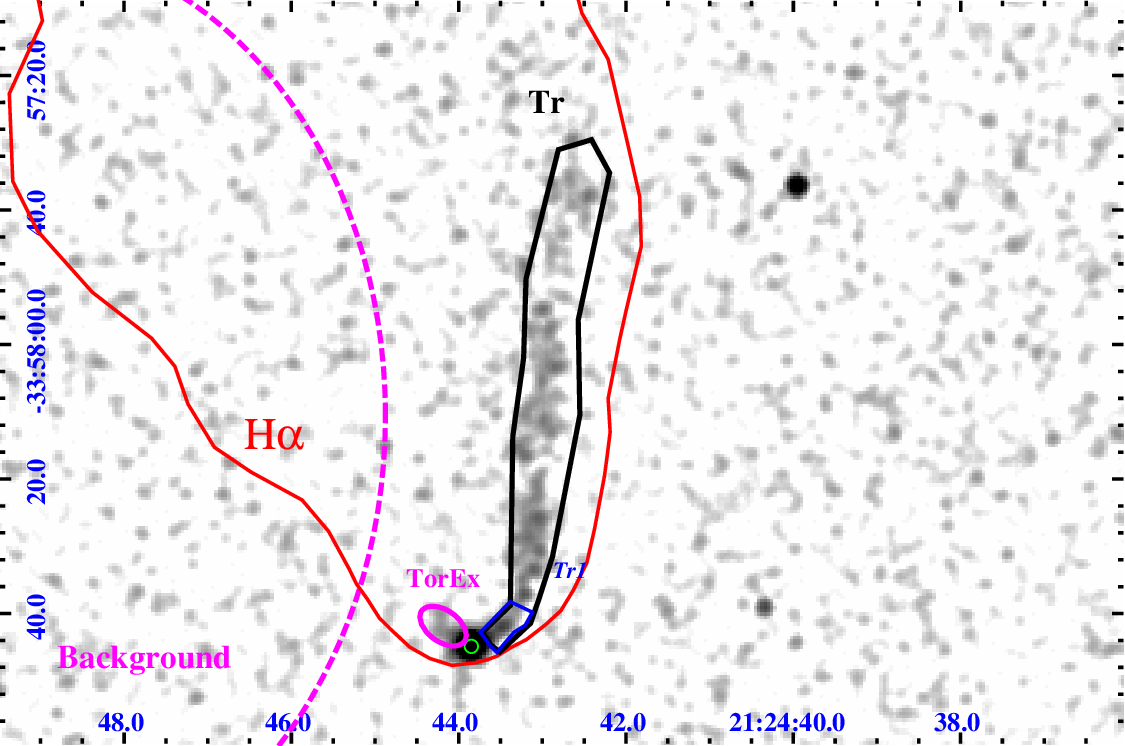}
\includegraphics{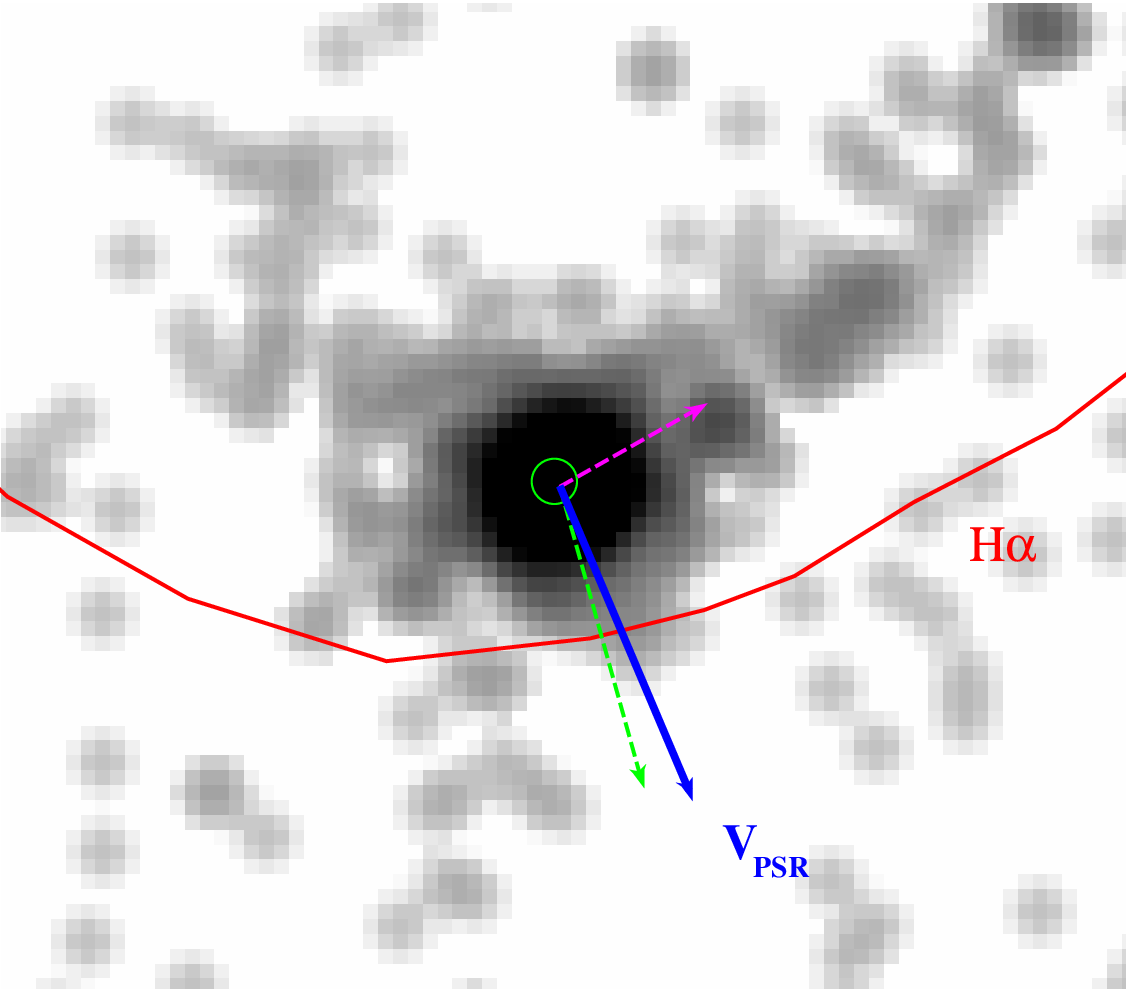}
\begin{center}
\caption{\label{CXO} 
Combined 0.3-3\,keV {\it CXO} image, smoothed with a $\sigma=1.5^{\prime\prime}$ Gaussian.
The limb of the $H\alpha$ bow shock encloses the trailed reverse shock X-ray PWN.
Some spectral extraction regions are indicated.
The $20^{\prime\prime}$ inset shows the region around the pulsar smoothed at $0.75^{\prime\prime}$,
with the H$\alpha$ limb location for reference. The proper motion vector (green dashed arrow) and
the motion corrected to the local standard of rest (blue arrow) are both plotted with a length
showing the displacement over 100\,y. The initial trail (polar outflow) axis is shown in magenta.
An equatorial extension to the PSF is visible normal to this initial axis.
}
\end{center}
\vskip -0.7truecm
\end{figure}

	We have made some basic spectral measurements of these X-ray structures.
Using the regions defined in figure 1, we extracted counts from the aligned exposures
and fit to thermal and power law models, with Galactic absorption using the \citet{wam00}
{\tt xstbabs} model. Background was extracted from a nearby source-free region on the S3 chip.

\begin{deluxetable}{lllllll}[t!!]
\tablecaption{\label{SpecFits} ACIS Spectral fits}
\tablehead{
\colhead{Reg.}&\colhead{${\rm N_H}$}&\colhead{$\Gamma$}&\colhead{$f_{PL}$}&\colhead{kT}&\colhead{$R_{BB}$ } &\colhead{P(Q)}\cr
\colhead{}&\colhead{b}&\colhead{}&\colhead{c} &\colhead{keV} & \colhead{km} & \colhead{}
}
\startdata
PS &$0.03_{-0.03}^{+0.5}$&  &   &$0.27_{-0.02}^{+0.03}$ &0.059 &$10^{-5}$\cr
PS &$20_{-3}^{+5}$       &$3.64_{-0.14}^{+0.15}$&$47._{-4}^{+5}$ & & &0.72\cr
PS$^d$ &$0_{-0}^{+9}$ &$2.6_{-2.3}^{+0.4}$& 0.039 & $0.25_{-0.06}^{+0.01}$& 0.078 &0.03\cr
PS &$0_{-0}^{+2}$ &$2.5_{-0.1}^{+0.2}$& $7_{-1}^{+1}$ & $0.25_{-0.01}^{+0.01}$& 0.047 &0.96\cr
Tr &$0_{-0}^{+5}$       &$2.15_{-0.14}^{+0.23}$&$1.8_{-0.2}^{+0.2}$ & & &0.81\cr
Tr &$1^a$       &$2.19_{-0.15}^{+0.15}$&$1.9_{-0.2}^{+0.2}$ & & &0.84\cr
TorEx &$1^a$       &$3.66_{-0.6}^{+0.7}$&$0.2_{-0.07}^{+0.07}$ & & &0.72\cr
Tr1&$1^a$       &$2.46_{-0.5}^{+0.5}$&$0.3_{-0.05}^{+0.05}$ & & &0.49\cr
Tr2&$1^a$       &$2.10_{-0.15}^{+0.16}$&$1.6_{-0.1}^{+0.1}$ & & &0.66\cr
\enddata
\tablenotetext{}{All errors are projected $1\sigma$ values. P(Q) gives the probability of the fit $\chi^2$.}
\tablenotetext{a}{ $N_H$ fixed at $10^{20} {\rm cm^{-2}}$.}
\tablenotetext{b}{ Wilms model absorption in $10^{20}{\rm cm^{-2}}$}
\tablenotetext{c}{ Unabsorbed 0.3-7keV flux in $10^{-14}{\rm erg\, cm^{-2} s^{-1}}$ .}
\tablenotetext{d}{ Double BB fit. `PL' parameters are $kT$ and $R_{BB}$ for the hot blackbody component.}
\end{deluxetable}
\bigskip

\begin{figure*}[t!!]
\vskip 5.5truecm
\includegraphics{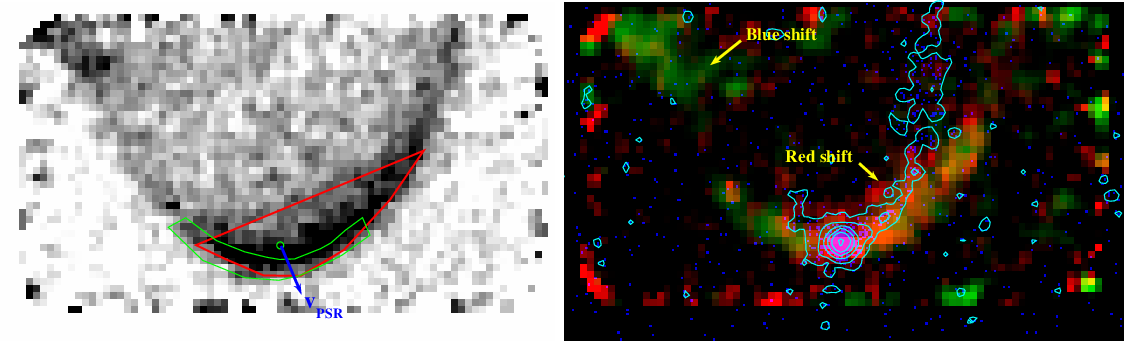}
\begin{center}
\caption{\label{KOALA} 
Left: KOALA field integrated over the H$\alpha$ line and background subtracted. The pulsar position 
and proper motion are indicated, along with the \citet{raet17} `F125LP' bow shock aperture (green)
and our `H$\alpha$ apex' aperture (red). Right: Apex velocity structure: KOALA maps
-69 to -37 ${\rm km\,s^{-1}}$ (green) and +48 to +77 ${\rm km\,s^{-1}}$ (red), with the {\it CXO}
0.3-3\,keV photons in blue, and cyan contours. The X-ray jet corresponds to a red-shifted 
portion of the H$\alpha$ shell, a blue-shifted region is seen to the NE.
}
\end{center}
\vskip -0.5truecm
\end{figure*}

	For the pulsar-dominated point source 
(PS, 2020 background-subtracted counts) we used a $1^{\prime\prime}$ radius aperture
to minimize contamination from the surrounding diffuse emission. Pile-up is negligible. 
A pure thermal fit indicates a low absorption, but high energy excess makes it statistically unacceptable.
A pure powerlaw fit is statistically adequate, but requires an unrealistically high absorption
to accommodate the low energy spectral peak. Thus a two component model is required. 
If two blackbodies are used the covariance is large and the parameters poorly determined.
The fit leaves residuals at high energy, leading to a low fit probability P(Q).
A powerlaw plus blackbody fit is fully adequate, with about half the counts from 
each component. The absorption is not well determined with an upper limit of $2\times 10^{20}
{\rm cm^{-2}}$. The thermal component has a high temperature $2.9\times 10^6$K
and a small effective area (spherical radius 47\,m, 2-D cap radius 94\,m at 410\,pc) 
indicating emission from 
a heated polar cap. We lack the sensitivity to meaningfully fit for a neutron star 
atmosphere model, but as seen in the XMM fits of \citet{z06}, a light element atmosphere
would decrease the cap temperature by $\sim 2\times$ and increase the cap radius by $\sim 4\times$. 
\citet{z06} also finds a pulse fraction of $56\pm14$\% in the {\it XMM} data (which includes the
extended emission). This indicates that the pulse fraction of the point source is very high. 
This pulsar is to be studied in the NICER core program and it is expected that a much more
detailed understanding of the pulse spectrum will be available soon. However the NICER steady
unpulsed component will include the extended non-thermal emission resolved in these ACIS data.

For the diffuse components, we first measured the spectrum of the full 
$80^{\prime\prime}$-length trail (Tr, 545 background-subtracted counts), finding a typical PWN powerlaw
index $\Gamma=2.15$. The absorption is only constrained to be $< 5 \times 10^{20} {\rm cm^{-2}}$.
To spectrally compare diffuse components we adopt a fixed $N_H=10^{20} {\rm cm^{-2}}$ (mid point
of the BB+PL $N_H$ range) and fit only for fluxes and power law indices. A check of the 
$\sim 8^{\prime\prime}$ trail before the `kink' where it approaches the H$\alpha$ limb
(Tr1, 64 background-subtracted counts), gave a somewhat softer index (2.5) than the remainder 
of the trail (Tr2=Tr-Tr1, 481 background-subtracted counts) beyond the kink (2.1). This difference is not
statistically significant, but we certainly do not detect synchrotron aging along
the trail. Interestingly the `torus' extension to the NE (TorEx, 42 background-subtracted counts) 
is substantially softer than the trail (at $\Gamma\approx 3.7$). 
The general diffuse emission around the point source seems similarly soft,
and this may bias the `pre-kink' trail index to larger values. The general situation
with a harder polar `jet' emission and softer equatorial `torus' emission seems
present in many PWNe. A good example is Geminga, where the long 
outer tails, which similarly bend to follow a bow shock shape while maintaining a hard
X-ray spectrum, are interpreted as polar jets \citep{poset17}.

\subsection{HST imaging}

	Given that the small shock stand-off is poorly resolved in ground-based H$\alpha$ imaging, we
attempted an observation of the H$\alpha$ limb using the {\it Hubble Space Telescope} WFC3 camera with
the narrow (1.4nm) F656N H$\alpha$ filter (Program 14364). Two orbits were used on August 14, 2016
with 5209\,s exposure in F656N and a 179\,s F606W exposures for continuum measurement. Despite careful
placement of the nebula apex near the readout node, pre-flashing of the WFC3 and tuned dithered
combinations of the long F656N frames, the nebula H$\alpha$ limb was not detected. We attribute this
to the poor WFC3 Charge Transfer Efficiency at the very low count level for this rather faint nebula. This
does mean that the H$\alpha$ limb is substantially resolved, as the count rate per pixel for a nebula
limb width $<0.3^{\prime\prime}$ should have allowed an WFC3 detection. Unfortunately the cycle 22
continuum images of Program 13783 (PI Pavlov) were not available while these observations were being
planned. These show a relatively bright and red pulsar point source, so that a larger fraction of the
orbit devoted to the F606W continuum frame would have given a red detection or useful upper limit.
In the end these {\it HST} exposures served to help reference the ground-based H$\alpha$ frames to
the precise pulsar position as measured in the F125LP and F475X continuum images.

\subsection{KOALA Spectroscopy}

	At present the best H$\alpha$ direct image available is the 600\,s SOAR/GHTS 
W012 image of BR14. We sought to improve this and to gather kinematic information on the 
forward shock structure by observing with the KOALA integral field unit (IFU) on the AAT under the NOAO
time exchange program (Project NOAO/36). KOALA was configured with
$1.25^{\prime\prime}$ sampling, so that the 1000 fibers covered 
$27.4^{\prime\prime} \times 50.6^{\prime\prime}$, sufficient to measure the PWN apex.
The fibers fed the AAOmega spectrograph with the red arm using the 2000R grating covering
$\sim 6273-6737$\AA~ at a resolution of 0.23\AA/pixel (10.5\,km/s at H$\alpha$). For the 
blue arm we used the 1500B grating, covering $\sim 4301-5077$\AA~ at $0.23$\AA/pixel 
(23.5\,km/s at H$\beta$). Observations were made on October 19-20, 2014 and on 
July 20, 2015. Seeing was variable, typically $1.5-2.0^{\prime\prime}$ but occasionally
as poor as $4^{\prime\prime}$. Both runs suffered
intermittent clouds and instrument problems. Nevertheless we were able to collect 
$13\times 1800$\,s on J2124, with the pointing dithered between exposures.
Flux calibration exposures of the continuum standards and of the emission lines
standards PN 205.8$-$26.7, PN211.0$-$03.5, PN253.3$-$03.9 and BoBn-1 were obtained with the same set-up. 
Since conditions were not photometric we will have non-statistical uncertainty
in the calibrated fluxes; we estimate these at $\sim 20$\% by comparing individual
exposures.

	All data were reduced with the AAT-provided 2dFdr-v6.2 software with KOALA configuration files.
Fibers were traced on the flat-field spectra, allowing optimal extractions of the arcs, standards
and objects. Wavelength calibrations were established against CuAr arcs while the spectra
were flat fielded using a combination of dome flats and twilight flats. Final flat fielding
was achieved by monitoring sky lines in the individual exposures. The 2dfdr software combines
the dithered exposures into data cubes sampled on a $0.75^{\prime\prime}$ grid. These were
sky subtracted by defining background regions well away from stars and nebulosity. These
regions were chosen to surround the nebula of interest, when possible. Residual flat field
errors led to some imperfection in the sky subtraction.  

The integrated H$\alpha$ image (Figure 2, left) shows an improved measurement of the apex surface brightness
distribution, although the IFU sampling and seeing substantially smooth the limb. Since no 
continuum stars appear in the KOALA field, we had to align the image using this blurred
H$\alpha$ limb, referenced through the SOAR H$\alpha$ image and the {\it HST} F606W frame.
This alignment is uncertain by $\sim 0.5^{''}$. J2124 is a relatively slow pulsar and so
the velocity spread of the bow shock is only modestly resolved, even for H$\alpha$.
Nevertheless we do see some interesting features. In the right panel of Figure 2, the redshifted
channel shows a ridge of emission inside the limb that follows the X-ray jet. There is also
a blue shifted region to the NE that appears to surround the pulsar position 500-600\,y in the past
and is presently marked by a bulge and brightening at the limb. Deeper velocity channel images with 
larger-scale coverage (and ideally few km/s resolution) would be needed to fully map these structures.

\begin{figure}[h!!]
\vskip 4.2truecm
\includegraphics{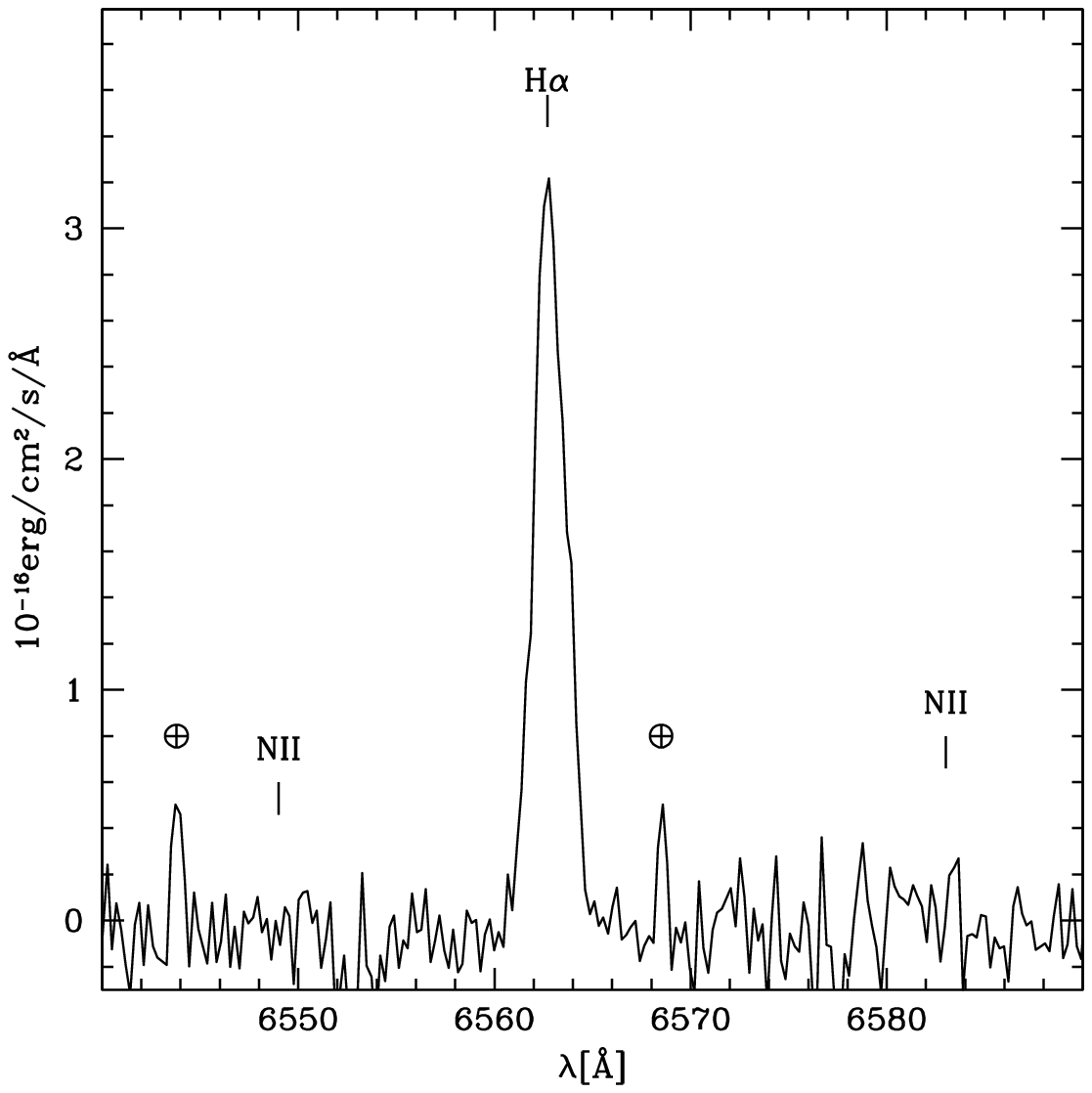}
\includegraphics{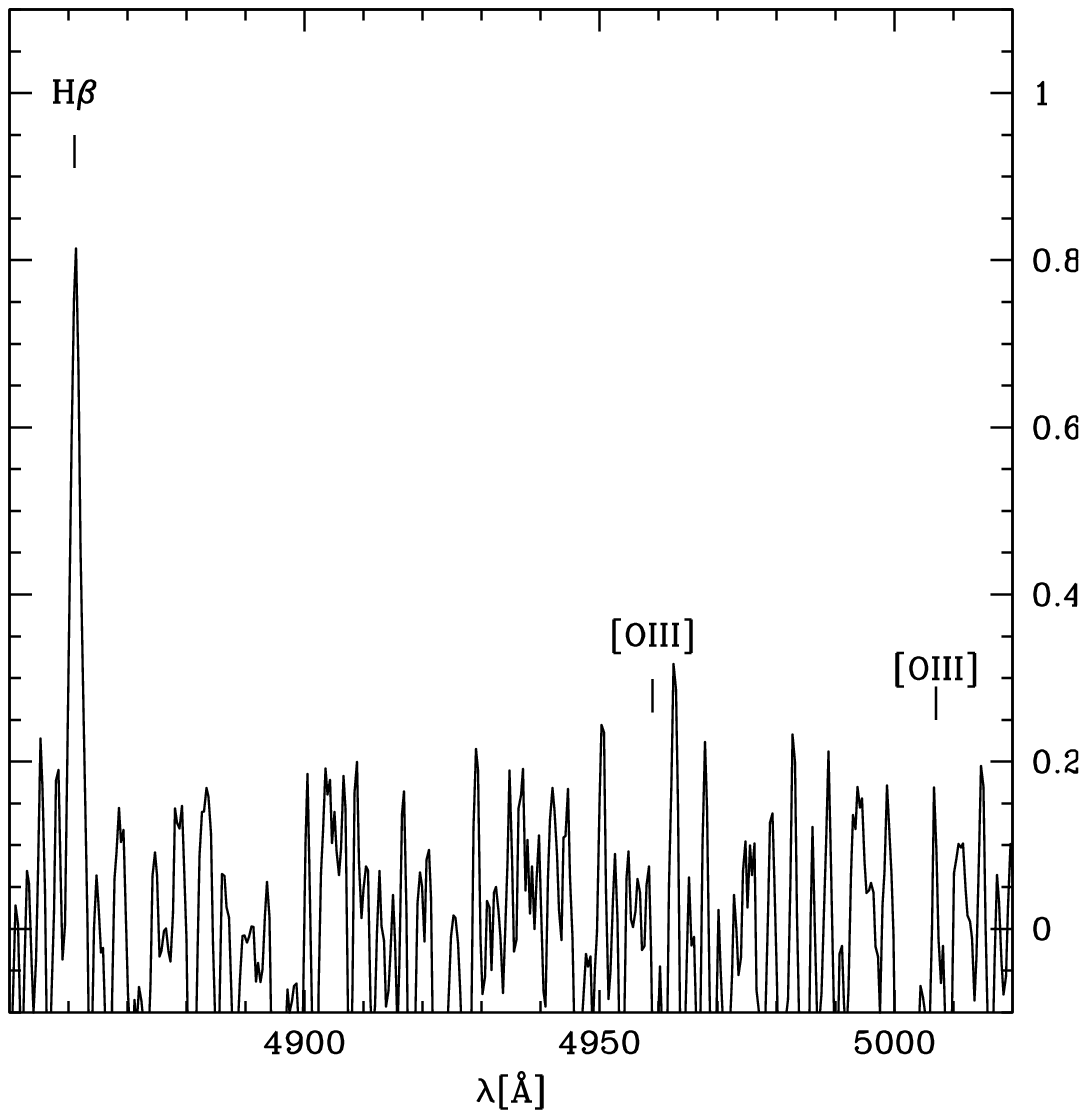}

\begin{center}
\caption{\label{KOALAspec} 
Portions of the KOALA IFU spectrum, integrated over the `H$\alpha$ apex' aperture and background
subtracted. Only the Balmer lines are well detected.
}
\end{center}
\vskip -0.5truecm
\end{figure}

	Our data provide an opportunity to check other emission lines and compare with shock models.
In \citet{raet17} a $52\,{\rm arcsec}^2$ aperture was defined covering the brightest portion of the
F125LP UV shock limb. This aperture does not correspond well to the H$\alpha$ limb. As confirmed
with the SOAR direct frame, portions of the UV flux lie ahead of this limb. Also, the overall shock
is distinctly wider to the west of the pulsar line of motion. We will be interested in the swept up 
ISM and the distribution about this pulsar velocity axis, including the front and
back sides. Thus we define a somewhat larger 
($108\,{\rm arcsec}^2$) aperture covering this `H$\alpha$ apex' and extending $3.5^{''}$ behind the
pulsar with a base perpendicular to the projected pulsar motion.

	Averaging over this aperture we measure the Balmer decrement 
$r_{\alpha\beta}=3.7\pm0.4$ with a negligible
correction for interstellar extinction (Figure 3). This is in reasonable agreement with the expected 
ratio $r_{\alpha\beta}=3.3$ at $v\approx 100 {\rm km \,s^{-1}}$ ($T\approx 10^5$K, Raga et al 2015).
The forbidden lines associated with radiative cooling are not detected, with upper limits
of $1\times 10^{-17} {\rm erg~cm^{-2}~s^{-1}}$ for OIII 5007 and 
$1\times 10^{-17} {\rm erg~cm^{-2}~s^{-1}}$ for NII 6582. \citet{raet17} present some radiative
models for the observed UV flux;
it is clear that the observed forbidden line/Balmer ratios are much lower than expected for
such a model; the bulk of the observed line emission comes from the non-radiative zone of
the bow shock.

\section{PWN/Bow Shock Structure}

	It is apparent from Figures 1 and 2 that the shock front is very close to the 
the pulsar along the proper motion axis and, as noted by BR14, the shock is very
flat at the apex and asymmetric about this velocity axis. In \citet{viget07} this
asymmetry was primarily ascribed to gradients in the ambient ISM. However, now the
ACIS image gives us a clear view of strong asymmetry in the (reverse shocked) pulsar wind.
Identifying the trail as a polar jet, as argued above, means that the counter-jet appears
at least $5\times$ fainter. This polar flux is also dramatically brighter than the
softer equatorial excess. This suggests a strongly asymmetric momentum distribution,
with a dramatic concentration to one side of the spin equator. We have extended standard
bow shock modeling to explore this possibility.


	In the X-ray the NW jet is strong, and well confined; the SE jet is faint or absent.
The softer outflow at PA $\sim 60^\circ$ which we identify as equatorial is also
more prominent on the trailing side. It is natural to ascribe this to the ram
pressure of the oncoming ISM. However we now have a good measurement of the pulsar-bow shock
standoff distance from the UV and the detailed limb and rough velocity structure
of the H$\alpha$ forward shock from the optical. Thus more detailed comparison with
shock model geometries is justified. 

	\citet{w00} provides elegant closed form expressions for the shape of the
momentum-balance contact discontinuity (CD). For an isotropic wind, the nose of the CD
has a characteristic standoff distance in the direction of the $v_p$ pulsar motion of
$$
r_0 = \left ( {\dot E}/4\pi \rho v_p^2 c \right ) ^{1/2},
\eqno(1)
$$
where we have assumed a relativistic massless wind ($\alpha=0$ in Wilkin's notation).

The formulae describe the CD shape for an axisymmetric momentum flux 
$p(\theta_\ast) = {\dot E}(\theta_\ast)/c = \Sigma_n c_n {\rm cos}^n \theta_\ast$, with 
${\dot E}(\theta_\ast )$ the co-latitudinal dependence of the wind and the spin axis at 
angle $\lambda$ to the space velocity with phase angle $\phi_\lambda$. See equation (A2) for
the relation of these angles to the velocity-aligned system.
\citet{w00} gives explicit expressions up to $n=2$.

Since the X-rays show the jet to be rather
narrow and since the H$\alpha$ apex is asymmetric about $v_{PSR}$, 
we found it helpful to expand further to better match the shock shape. 
The needed components $G_{w,{\tilde \omega}}$ and $G_{w,z}$ of the momentum integral 
are given to $n=4$ in the appendix. With these the shock limb is given as 
$$
r(\theta,\phi) = r_0 \left ( 6 G_{w,z} - 6 G_{w,{\tilde \omega}} {\rm cot} \theta \right )^{1/2}
/{\rm sin}\theta
\eqno(2)
$$
and the velocity of well-mixed flow tangent to this limb has amplitude
$$
v_t = v_{\rm PSR} 
[4G_{w,{\tilde \omega}}^2 + (2 G_{w,z} - {\tilde r}^2)^2 ]^{1/2}/{\tilde r}^2
\eqno(3)
$$
with ${\tilde r}=(r/r_0){\rm sin}\theta$.

	In the numerical realizations for a given set of $c_n$, $\lambda$ 
we compute the CD shape (Eq. 2) referenced to the proper motion axis ${\vec v}$.
We incline this vector by angle $i$, set $\phi_\lambda$ to orient the mis-aligned
spin axis relative to the plane of the sky and then compute the image by shooting rays through
this structure, summing the emissivity at each limb intersection and projecting
to the plane of the sky. The emissivity prescription is at present a simple 
scaling to the swept up ISM density at $r(\theta, \phi)$. We also take the amplitude
of the velocity in Equation 3, compute the tangent vector in the constant $\phi$ 
plane at $r(\theta, \phi)$ and
project this velocity along the Earth line-of-sight. The surface flux is assigned
to the velocity vector representing this tangent flow (and an additional component in
a narrow distribution around zero velocity to represent emission from the collisionally
excited neutrals in the shocked ISM) and summed into velocity planes of a data cube.
The entire structure is rotated to match the observed  ${\vec v}$ axis and a world
coordinate system is assigned to match the pulsar position and image scale.
While the intensity distribution is arbitrary and will not match the details of
the observed bow shock, the CD shape (and approximate velocity distribution)
can be compared with the KOALA data cubes.

	We should recall that the observed radiation comes from the forward (Balmer, UV) and 
reverse (X-ray) shocks. The forward shock stands off from the CD, by a factor estimated
as $\sim 1.3\times$ at the nose of a bow shock with a simple isotropic wind \citep{arc92,b02}.
The shock pressure serves also to
make the forward shock limb smoother, so it is less sensitive to the fine details of the
wind anisotropy. Another difference to the simple model is that some line emission
will be prompt, as the neutral ISM particles enter the shocked ISM and experience excitation
and charge exchange into excited states. Only after the medium is largely ionized will
$v_t$ fully describe the flow.

	The jet asymmetry can be reproduced using $c_1$ and/or $c_3$ while the small
standoff and jet dominance require large $c_2$ and/or $c_4$. Our goal is to match the
overall shape of the H$\alpha$ limb, recognizing that the post-shock pressure will make
this larger and smoother than the computed CD surface. Any attempt to match the
shape with a dominant equatorial flow (large negative $c_2$) and the spin axis aligned with
the proper motion fails, given the E-W asymmetry of the limb about the proper
motion axis. Indeed, we were not able to make a sufficiently `flat' bow shock 
even for highly equatorial models viewed edge-on (spin axis in the plane of the sky); their
apex curvature was always too large. Adding $c_4$ terms did not significantly help.

	Instead, the most promising solutions invoke polar (jet-dominated) models,
with a large ${\vec \Omega}-{\vec v}$ angle. This accords with our spectral inference
that the trail is a polar jet, and additionally allows us to produce E-W asymmetry
by using the $c_1$ and $c_3$ terms. In our modeling we fix the proper motion at the observed
$203^\circ$ angle, with ${\vec v}$ inclined by an angle $i$ to the line of
sight. We match the projected spin axis ${\vec \Omega}$ with the $PA\approx -60^\circ$ axis 
of the base of the observed jet. Thus this imposes a constraint on the ${\vec \Omega}-{\vec v}$
angles $\lambda$ and $\phi_\lambda$. The shape is controlled by the four parameters $c_1 - c_4$,
and an overall normalization factor, which we report for convenience as $\theta_0$,
the standoff angle of the isotropic wind with the same total momentum flux. This normalization
could also be quoted via $c_0$ and a rescaling of the other coefficients. This is a total
of seven model parameters. Unfortunately, the finite pressure standoff means that we cannot 
directly fit this CD sum to the shock limb shape. But matching the overall shape is surprisingly
constraining and the parameters are fairly well determined in the context of this model.
Full MHD simulations will be needed for a detailed match.

\begin{figure}[t!!]
\vskip 4.2truecm
\includegraphics{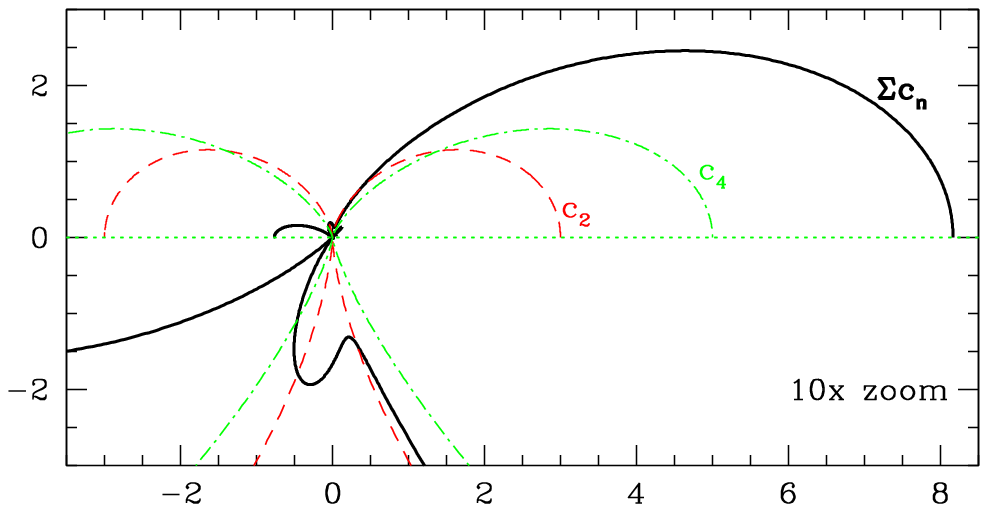}
\begin{center}
\caption{\label{Power} 
Polar plot of PWN momentum distribution for our J2124 model. Symmetric ${\rm cos}^2\theta$ (red)
and ${\rm cos}^4\theta$ (green) winds with the same integrated power are shown for comparison. The
bottom panel magnifies the arbitrary scale $10\times$ to show the weak quasi-equatorial component.
}
\end{center}
\vskip -0.5truecm
\end{figure}

	We have computed a range of CD models, adjusting the $c_n$ and angles, while
holding the velocity vector fixed at the corrected proper motion direction and
the spin axis fixed along the jet base. Figures 4 and 5 show the best match found. This is 
a model with the pulsar velocity inclined $i=120^\circ$ from the line of sight ($30^\circ$
into the plane of the sky), with the spin axis at $\lambda=95^\circ$ to ${\vec v}$,
directed at $\phi_\lambda=10^\circ$. The momentum coefficients are $c_1=-0.3$,  $c_2=-0.2$,
$c_3=4.0$ and $c_4=4.5$, and the isotropic equivalent standoff angle is $\theta_0=2.1^{\prime\prime}$.
Figure 4 shows the polar plot of the momentum distribution, while figure 5 compares
directly with the H$\alpha$ limb. For the left panel of figure 5 we have
expanded to $1.3\times \theta_0$ to show the shape match to the limb. Note that for these
coefficients the momentum flux has a weak near-equatorial (`torus') component in Figure 4 at 
$\theta \approx 110^\circ$ and a near-null at $\theta \approx 120^\circ$. These show up
in the contact discontinuity shape as the small bulge near the nose and the edge-brightened
indentation just to its east (Figure 5). While a momentum bulge from the weak torus is physically 
plausible, the near-null is likely just an artifact of the limited number of $c_n$. Neither CD feature is
expected to show in the forward shock (H$\alpha$) structure, being smoothed by the shocked ISM
pressure. The model's projected spin axis lies
along the {\it CXO} jet emission and the momentum for the stronger forward jet into the
plane of the sky makes a PWN bulge in that direction. The red-shifted KOALA velocity channels
are concentrated along this bulge with smaller excursions for the torus' momentum into the
plane of the sky and for the weak counter-jet (cf. Figure 2, right).

	Since we can only match the observed shape, rather than fit, we cannot quote formal
errors. However, the model shape departs noticeably from the observed limb shape 
when we shift the parameters by
$\sigma_i \approx 20^\circ$,
$\sigma_\lambda \approx 10^\circ$,
$\sigma_{\phi_\lambda} \approx 10^\circ$ (linked to $\lambda$ to set the projected PA),
$\sigma_{\theta_0} \approx 0.1^{\prime\prime}$,
$\sigma_{c_1} \approx 0.2$,
$\sigma_{c_2} \approx 0.1$,
$\sigma_{c_3} \approx 0.3$ and
$\sigma_{c_4} \approx 0.2$.
Figure 4 shows that even a 4$^{th}$ order expansion does not produce a jet as collimated as
that seen in the ACIS image. For this one needs to go to $n=8-10$ and a numerical approach
will be more productive. This may help to further flatten the shock apex and tune the fine
structure of the torus and counter-jet contributions.


\begin{figure*}[t!!]
\vskip 3.8truecm
\includegraphics{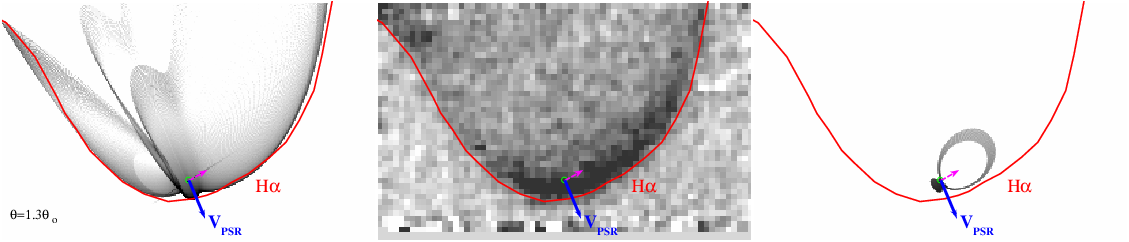}
\begin{center}
\caption{\label{Power} 
PSR J2124$-$3358 CD model. Left: Direct image of model, with CD scale amplified $1.3\times$ to show the limb
match. Arrows from the pulsar point source show the spin and jet axes. Note that the groove in 
the model surface is a result of the equatorial component and near-null (see fig 4); such structure 
is an artifact of the limited number of $c_n$ used and should not appear in the H$\alpha$ limb
of the pressure-smoothed forward shock. Middle: velocity-integrated KOALA image,
with the pulsar velocity and X-ray jet axes indicated. Right: a red-shifted model channel showing
$+55-65 {\rm km\,s^{-1}}$ emission. The data have a similar, but narrower, red extension at 
along the jet PA (Figure 2).  The H$\alpha$ limb is shown in red in the three panels, for comparison.
}
\end{center}
\vskip -0.5truecm
\end{figure*}

\section{Models and Masses}

	The model in section 3 allows us to estimate the total energy flux of
PSR J2124$-$3358. Because many kinematic parameters are available from the precision
pulsar timing, this becomes an important test of the pulsar physics. The basic argument
was presented in BR14: with a known efficiency of Balmer photons/HI atom entering
the shock, the observed H$\alpha$ flux plus the bow shock geometry give the upstream neutral
density.  With a distance and proper motion and a possible correction for ionization fraction, 
this also gives the incident momentum flux. Balancing this with the PWN flux, including 
correction for the anisotropy, we obtain the spindown power and, using the timing parameters, 
the neutron star moment of inertia.

	Transverse to the projected pulsar motion the H$\alpha$ apex aperture extends
$26.6^{\prime\prime}$ in the sky plane. Our asymmetric shock model indicates extension along the jet
axis. The 3-D apex aperture has a nearly elliptical cross section with axes $26.6^{\prime\prime}
\times 13^{\prime\prime}$. With a space velocity 
$\mu_\perp d/{\rm sin}i=127 d_{410}({\rm sin}i/0.866)^{-1}{\rm km\, s^{-1}}$
(for the corrected transverse velocity and a distance scaled to 410\,pc) 
the apex zone sweeps out a volume $1.3 \times 10^{41} d_{410}^3({\rm sin}i/0.866)^{-1}{\rm cm^3s^{-1}}$.
In the KOALA data cube the observed $H\alpha$ photon rate from this zone is 
$2.3\times 10^{-4} {\rm H\alpha \, cm^{-2} s^{-1}}$ while in the direct SOAR image we obtain
$2.9\times 10^{-4} {\rm H\alpha \, cm^{-2} s^{-1}}$. We will adopt the average,
$f_{H\alpha}=2.6\times 10^{-4} {\rm H\alpha \, cm^{-2} s^{-1}}$, although the 
transparency issues during the KOALA
run suggest that this is conservative. This flux corresponds to a nebula apex luminosity
of $5.2 \times 10^{39} d_{410}^2 {\rm H\alpha \,s^{-1}}$. If we assume electron-ion equilibration
the H$\alpha$ yield is $\epsilon_{\rm H\alpha} =0.6 (v/100 {\rm km\,s^{-1}} )^{-1/2}$ per 
neutral crossing the shock \citep{hm07}, giving $9.8 \times 10^{39} d_{410}^{5/2} 
({\rm sin} i/0.866)^{-1/2}$HI/s.  This implies an upstream HI density 
$n_{HI} =0.076 d_{410}^{-1/2}({\rm sin} i/0.866)^{1/2}{\rm cm^{-3}}$,
with a total density $(1-x_i)^{-1} \times$ larger, for an ionization fraction $x_i$.  
We will assume electron-ion equilibrium in 
the following sums, as the disequilibrium H$\alpha$ efficiency is nearly $10\times$ smaller at 
these velocities, leading to very large densities and an unreasonable spindown power.

	This $n_{HI}$ is somewhat unusual. By volume most of the HI is in the Warm Neutral
Medium (WNM), which at J2124's 290\,pc height above the plane has a typical density
$n_{WNM} \approx 0.25 \,{\rm cm^{-3}}$ and is largely neutral with $x_i\approx 0.05$. This
would produce $n_{HI}$ and nebular H$\alpha$ flux $\sim 3\times$ larger than observed. The Warm Ionized
Medium (WIM), in pressure equilibrium at $n_{WIM}\approx 0.125 \,{\rm cm^{-3}}$ and $x_i\approx 0.95$,
has a much lower HI density and is not a viable alternative. So we infer that the local medium
is partly ionized and has a proton density larger than that of the observed HI. If the ionization is
general, preserving pressure equilibrium with the WNM/WIM, the observed $n_{HI}$ implies $x_i=0.55$.
If the ionization is local so that the upstream medium can be over-pressured, the 
typical $n_{WNM} \approx 0.25 \,{\rm cm^{-3}}$ density gives $x_i=0.70$.

	Our model determines the isotropic equivalent standoff angle as $2.1^{\prime\prime}$, i.e.
$r_0=1.29 \times 10^{16} d_{410}$cm. Thus from equation (1) we write
$$
{\dot E} = 4\pi^2 I {\dot P}/P^3 = 4\pi \rho v^2 c r_0^2.
\eqno (4)
$$
The observed flux $f_{H\alpha}$ gives us our estimate of $n_{HI}$ which we convert to mass density as
$\rho= \mu m_p n_{HI}/(1-x_i)$ for mean mass per particle $\mu m_p$. Plugging in gives
${\dot E} = 1.77 \times 10^{33} d_{410}^{7/2}({\rm sin} i/0.866)^{-3/2} (1-x_i)^{-1} {\rm erg\,s^{-1}}$.
Applying the Shklovskii correction \citep{shk70}, this gives us the neutron star moment of inertia as
$$
I = {\dot E} P^3/[4\pi^2 ({\dot P}_{obs} - v_\perp^2P/dc)]
\eqno (5)
$$
For J2124 we have $P=4.93$ms, ${\dot P}_{obs}=2.057\times 10^{-20}$ 
(Shklovskii-corrected ${\dot P}=7.27\times 10^{-21}$), giving
$$
I_{45}=0.74 d_{410}^{7/2} ({\rm sin} i/0.866)^{-3/2} (1-x_i)^{-1}
\eqno (6)
$$
in standard units of $10^{45} {\rm g\,cm^2}$. From the ionization discussion above we see that this
is $I_{45} = 1.66$ for ionization maintaining WNM pressure equilibrium ($x_i=0.55$) and
$I_{45} = 2.43$ if we partly ionize the typical WNM density ($x_i=0.70$).

	We would like to relate this moment of inertia to the neutron star mass. Given
the existence of $2M_\odot$ neutron stars, the equation of state is fairly stiff and 
from \citet{ls05} we see that 
$$
I_{45} \approx [0.8-1.2] (M/M_\odot)^{3/2},
\eqno (7)
$$
where the prefactor $[\,]$ covers the range of acceptable EOS from relatively soft to very stiff. 
This can be used to make a neutron star mass estimate
\setcounter{equation}{7}
\begin{equation}
\begin{split}
M_{NS}= & [0.73-0.95] \left ( f_{H\alpha} \over {2.6 \times 10^{-4} {\rm cm^2s^{-1}}} \right )^{2/3}
\\
 & (1-x_i)^{-2/3} d_{410}^{7/3} ({\rm sin} i/0.866)^{-1}M_\odot.
\end{split}
\end{equation}
This estimate again supports our inference of substantial ionization of the ambient WNM,
but highlights the uncertainty introduced by this factor. For $x_i=0.55$ the allowed range 
is $1.24-1.63 M_\odot$, allowing but not demanding substantial accretion during spin-up.
For $x_i=0.70$ we infer $1.60-2.10 M_\odot$, so that significant mass growth is required.
However, even then uncertainty in the other parameters precludes strong conclusions.
In particular while the present parallax error is only -17\%/+22\%, the strong distance
dependence leads to a -35\%/+59\% uncertainty in the neutron star mass. Systematic (model) uncertainties
are at present smaller; the crucial factors are the nebula cross section
and pulsar velocity inclination, which are constrained by the direct imaging. In the context
of the model matches, our `chi-by-eye' approach shows a $\sim 10\%$ range in 
$I_{45}$, varying these geometric terms over the range of acceptable fits. We consider this 
a `systematic' error range for this model -- of course if the model is incorrect the true
$I_{45}$ could differ by more.  Uncertainty in the measured $f_{H\alpha}$ contributes
$\sim 20\%$ to $I_{45}$ or $\sim 13\%$ to $M_{NS}$.

\section{Conclusions}

	We measure strong asymmetry in PSR J2124$-$3358's bowshock and PWN.
In the ACIS X-ray data,
the reverse shock emission seems to be dominated by a polar jet trailing the pulsar
motion. There are a few photons plausibly assigned to counter-jet emission, but comparing
similar areas in the few arcsec at the base of the jets (before the bow shock) the counter-jet
is at least $5\times$ fainter. There is also a weak equatorial outflow, 
with a softer spectrum. Both polar and equatorial emission appear swept back
by the pulsar motion and bounded by the UV/H$\alpha$ PWN.

	The {\it HST}-detected UV bow shock lies near the forward edge of the observed
H$\alpha$ emission. This adds credence to the idea that the observed F125LP flux arises in the
shocked ISM. However, our KOALA IFU spectroscopy of the bow shock apex shows that the
cooling lines from such shocked emission do not dominate the nebula. For example, 
\citet{raet17} Figure 9 shows OIII 5007/H$\beta \approx 1$ for the thermal emission from
a cooling plane shock; our
upper limit is 5007/H$\beta < 0.06$. Also from Figure 6 of \citet{raet16} we infer
a predicted thermal plane shock H$\alpha$ flux of $\sim 3.3 \times 10^7 {\rm H\alpha\,cm^{-2}s^{-1}}$
at the source or $2.8\times 10^{41} {\rm H\alpha\,s^{-1}}$ for our full apex zone. Our 
observed flux for this bow shock zone
is only $2\%$ of this value (of which only $\sim 5$\% can be thermal emission). Clearly
radiative emission can only play a small role in the bow shock, where the shocked gas
is advected far down stream before it cools. This explains the extreme Balmer dominance and 
indicates that we should also attribute the UV emission to a non-thermal model, in which 
PWN $e^\pm$ infiltrate the shocked ISM and radiate. We know of no detailed model for this case,
but multi-band continuum imaging of the shock should be able to distinguish the synchrotron
emission from the thermal scenario.

	The KOALA data also provide an improved S/N map of the nebula limb (albeit
at low spatial resolution) and information on the velocity-dependent morphology. 
Upgrading the analytic bow shock models to allow narrower polar outflows, we have attempted to
match the KOALA shape and velocity dependence, while holding the pulsar transverse
velocity vector and the projected spin (jet) axis vector fixed. The result indicates
a pulsar moving $\sim 30^\circ$ into the plane of the sky with an approximately orthogonal
spin axis. The very large jet/counter-jet asymmetry (a factor of $\sim 10$ for the integrated flux
in our model) induces an E-W asymmetry about the proper motion axis and makes the bowshock
ovoid in its cross section.

	The origin of this asymmetry is unclear. Images of other PWNe with apparent polar jets
do show asymmetries, but these are generally smaller \citep[e.g.][]{karet17}. At the base of 
such jets unequal fluxes can be attributed to Doppler beaming but on larger scales sweepback
is the likely culprit. In J2124 we might have attributed the dimness of the 
counter-jet to ram pressure suppression or sweepback, but we have the evidence of the H$\alpha$/UV forward
shock shape to show that the standoff (and hence the outflow/ram pressure ratio) is smaller on
that side of the spin axis. Hence we infer a true momentum asymmetry. Modest momentum
asymmetries have long been inferred when the pulsar has an off-center dipole field, the
`Harrison-Tademaru' effect \citep[see][]{lcc01}. For a simple vacuum dipole, the momentum
asymmetry is, however, $<2\times$; we can only speculate that a more complex multipole
configuration or plasma effects might increase the allowed asymmetry.

	In the context of this asymmetric model, the observed H$\alpha$ flux and nebula geometry
give an estimate of the PWN momentum flux needed to balance the oncoming ISM. This momentum
flux gives the spindown power and via the pulsar timing parameters an estimate of the
moment of inertia. These estimates depend on the upstream ionization, which we estimate as 
$x_i=0.55-0.70$. The PWN itself may provide a source of such ionization. We checked if the
UV flux (dominated by the bow shock) itself could photoionize the upstream WNM.
The observed $L_{\rm 6.2eV-9.9eV} = 1.9 \times 10^{29} {\rm erg\, s^{-1}}$ \citep{raet17}
provides an H-ionizing cross-section flux of
$a {\dot N}_i = \int dN/d\epsilon a_H(\epsilon) d\epsilon \approx 4 \times 10^{22} {\rm cm^2 s^{-1}}$
(for an assumed flat spectrum), which only ionizes over a characteristic distance
$r_i = a {\dot N}_i/4\pi v$ \citep{vkk01} or $\theta_i \approx 0.04^{\prime\prime}$ for the J2124
distance and parameters. Since the bow shock is detected in neither the X-ray nor the optical,
the index and flux cannot be very different than assumed here; with $\theta_i \ll$ the standoff angle
$\theta_0$, the UV cannot strongly pre-ionize the ambient medium.

	However the UV-emitting electrons themselves are a possible source of cosmic rays
to pre-ionize the medium. We already infer that the PWN $e^\pm$ can cross the CD to generate
UV emission in the forward bow shock.
Since the pulsar spindown power goes mostly into $e^\pm$ the energy density at offset
$\theta$ is $\sim {\dot E}/(4\pi d^2\theta^2 c) \approx 100/\theta_{\rm arcsec}^2 {\rm eV \, cm^{-3}}$,
so if a few percent of the PWN power escapes it can dominate the local ionizing cosmic ray 
density. Such escape may be particularly
enhanced for J2124 because of the PWN anisotropy and the small standoff at the apex. Recall
that the Eastern (E) section of the shock apex, where the counter-jet impacts and the UV flux (and 
hence ISM penetration of PWN $e^\pm$) is largest, also shows fainter H$\alpha$ emission
(Figure 2). One can speculate that this is partly due to greater pre-ionization in this
portion of the bow shock. However one should also note that the faintest portion of the E limb 
is nearly parallel to the proper motion: very little fresh HI enters the shock in this region 
in any event.  One might further speculate that if, contrary to expectation, very little ionization
occurs in the bow-shocked ISM \citep{mlv15}, then the effective (ionized) ISM density might
be much larger in the E half of J2124's bow shock than in the west. Such non-uniform ionization 
(traceable to the PWN outflow geometry) might then contribute to the asymmetry of the
H$\alpha$ limb. We do not pursue this further here, although it is a fruitful scenario for 
additional study.

	The WNM ionization, while almost certainly present, is thus poorly understood.
In turn this leads to uncertainty in the J2124 mass and moment of inertia. Accordingly bow shocks in
lightly ionized $x_i \approx 0$ WNM will provide the most robust evidence for large $I_{45}$ and
$M_{NS}$ since the ionization correction is small. This appears not to be the case for J2124.
If we can observationally constrain the upstream temperature and ionization state this would help reduce
the uncertainty. One approach would be to obtain high resolution spectra measuring
the narrow (collisional excitation) line component in the non-radiative shock, whose width is
sensitive to the upstream pre-ionization and heating. This component is not 
detected in our moderate resolution spectra. However at present other factors, principally the distance
uncertainty, dominate the $I_{45}$ error budget, so better distances and H$\alpha$ flux
measurements are needed before ionization and model uncertainties are the limiting factor.

	We are left with an interesting, but rather imprecise result. With our best
estimate of the state of the upstream medium (WNM with preshock ionization) we infer
$M_{NS} = 1.60-2.10 M_\odot$ (depending on EOS choice), which implies substantial
mass accretion during the recycling. If the pre-ionization only produces pressure equilibrium
our measured $n_{HI}$ leads to an estimate $M_{NS} = 1.24-1.63 M_\odot$, so softer
allowed equations of state imply mass growth during recycling, while stiffer EOS do not
require it. Of course, with a combined error as large as 50\%, dominated at present
by the parallax uncertainty, none of these conclusions are very robust. Still, we have
used our PWN measurements to gain some insights into the neutron star spin orientation
and some spindown power constraints. It will be particularly interesting to compare
the results of our modeling with measurements of spin orientation and $M/R$ constraints
from the new NICER mission where PSR J2124$-$3358 is one of the 3-4 prime early 
targets \citep{oet16}. Indeed, if NICER succeeds in making a radius measurement for 
this pulsar then our $I$ estimate will be particularly interesting, since no orbital 
mass measurement is available and only light bending constraints are expected. 
\bigskip
\bigskip

We thank Newton C.-S. Cheng for assistance with reduction of the KOALA data cube and the anonymous
referee whose careful reading caught a number of typographical errors in the equations and whose 
comments led to improvement of the text.
This work was supported in part by NASA grants GO6-17059X (CfA) and GO-14364 (STScI).

\appendix
\section{Bow Shock Momentum Functions}

	For the convenience of others wishing to simulate asymmetric thin bow shocks, we provide
a further expansion of the momentum integrals in \citet{w00}. These expressions assume
an axisymmetric relativistic massless pulsar wind (Wilkin's $\alpha=0$) with momentum distribution
\begin{equation}
p (\theta_\ast ) = \sum_n c_n {\rm cos}^n \theta_\ast
\end{equation}
with $\theta_\ast$ the co-latitude to the pulsar spin axis. The relationship between
the spin- ($\theta_\ast$, $\phi_\ast$) and velocity- ($\theta$, $\phi$) referenced coordinate
systems is
\begin{equation}
\begin{split}
{\rm sin}\theta_\ast {\rm cos} \phi_\ast = \, & {\rm sin}\theta\,  {\rm cos} \phi  \\
{\rm sin}\theta_\ast {\rm sin} \phi_\ast = \, & {\rm sin}\theta\,  {\rm sin} \phi\, {\rm cos}\lambda - {\rm cos}\theta \, {\rm sin} \phi \\
{\rm cos}\theta_\ast = \, & {\rm sin}\theta\, {\rm sin} \phi\,  {\rm sin}\lambda - {\rm cos}\theta \, {\rm cos} \phi
\end{split}
\end{equation}
where, following Wilkin's notation the pulsar spin axis is oriented at angle $\lambda$ 
to the pulsar velocity vector with orientation angle $\phi_\lambda$; $\phi_\lambda=0$ places
the spin axis in the plane of the sky. 

In writing the $G$ integrals we define $p={\rm sin}\phi \,{\rm sin} \lambda$ and $q={\rm cos} \lambda$
for convenience.  We will carry out the expansion to $n=4$
so the normalization condition is $c_0=1-c_2/3 -c_4/5$. Thus $c_{2,4}$ determine the equatorial
concentration, while $c_{1,3}$ determine the polar asymmetry.  
Then at angle $\theta$ to ${\vec v}_{PSR}$ the parallel component is
\begin{equation}
\begin{split}
G_{w,{\tilde \omega}} = &
(c_0/2) \, [\theta-{\rm cos}\theta {\rm sin}\theta] + \\ 
& (c_1/3)\,[4p (2+{\rm cos}\theta) {\rm sin}^4(\theta/2)  + q {\rm sin}^3\theta] + \\
& (c_2/32)\,[q^2(4\theta-{\rm sin}4\theta)+ p^2(12\theta -8 {\rm sin}2\theta + {\rm sin}4\theta)  
   + 16pq{\rm sin}^4\theta] + \\
& (c_3/30)\,[ 3 pq^2 (4+{\rm cos}^3\theta[3{\rm cos}2\theta -7]) + 
   8p^3 (19+18{\rm cos}\theta +3 {\rm cos}2\theta) {\rm sin}^6[\theta/2] +
   q^3 (7+3{\rm cos}2\theta){\rm sin}^3\theta +18 p^2q{\rm sin}^5\theta] + \\
 & (c_4/192)\, [64 pq^3(2+{\rm cos}2\theta){\rm sin}^4\theta +
    128p^3q {\rm sin}^6\theta + 
    q^4(12\theta+3{\rm sin}2\theta -3{\rm sin}4\theta - {\rm sin}6\theta) + \\
& \qquad\qquad\qquad    p^4(60\theta-45{\rm sin}2\theta +9{\rm sin}4\theta - {\rm sin}6\theta) +
    6p^2q^2 (12\theta-3{\rm sin}2\theta -3{\rm sin}4\theta + {\rm sin}6\theta)]. 
\end{split}
\end{equation}
The component perpendicular to the velocity is
\begin{equation}
\begin{split}
G_{w,z}= &
(c_0/2) \,[{\rm sin}^2 \theta ] + \\
& (c_1/3)\,[q (1-{\rm cos}^3\theta)+p {\rm sin}^3\theta] + \\
& (c_2/16)\,[ 4 q^2 (1-{\rm cos}^4\theta) +4 p^2 {\rm sin}^4\theta + p q (4\theta - {\rm sin}4 \theta  )] + \\
& (c_3/10)\,[ 2 q^3 (1-{\rm cos}^5\theta) + p^2q (4+{\rm cos}^3\theta [3 {\rm cos}2\theta -7]) +
    p q^2 (7+3{\rm cos}2\theta) {\rm sin}^3\theta +2 p^3 {\rm sin}^5\theta] + \\
& (c_4/48)\, [8 q^4(1-{\rm cos}^6\theta)+24p^2q^2 (2+{\rm cos}2\theta) {\rm sin}^4\theta +
    8 p^4 {\rm sin}^6\theta + 
    pq^3 (12\theta+3{\rm sin}2\theta -3{\rm sin}4\theta - {\rm sin}6\theta) + \\
& \qquad\qquad\qquad    p^3q (12\theta-3{\rm sin}2\theta -3{\rm sin}4\theta + {\rm sin}6\theta)]. 
\end{split}
\end{equation}

\end{document}